\documentclass[prb,twocolumn,showpacs,floatfix]{revtex4-1}
\usepackage{graphics,epsfig,amsfonts,amssymb,amsmath,soul}
\usepackage[T1]{fontenc}
\usepackage[utf8]{inputenc}
\usepackage{lmodern}
\usepackage[english]{babel}
\usepackage{wasysym}
\usepackage{color}
\begin{document}
\selectlanguage{english}
\title{Bursty magnetic friction between polycrystalline thin films with domain walls}

\author{Ilari Rissanen$^1$}
\email{ilari.rissanen@aalto.fi}
\author{Lasse Laurson$^2$}

\affiliation{$^1$Helsinki Institute of Physics and Department of Applied Physics, 
Aalto University, P.O. Box 11100, FI-00076 Aalto, Espoo, Finland}
\affiliation{$^2$Computational Physics Laboratory, Tampere University, 
P.O. Box 692, FI-33014 Tampere, Finland}

\begin{abstract}
\noindent Two magnets in relative motion interact through their dipolar fields, 
making individual magnetic moments dynamically adapt to the changes in the energy 
landscape and bringing about collective magnetization dynamics. Some of the  
energy of the system is irrevocably lost through various coupling mechanisms between 
the spin degrees of freedom and those of the underlying lattice, resulting in magnetic 
friction. In this work, we use micromagnetic simulations to study magnetic friction 
in a system of two thin ferromagnetic films containing quenched disorder mimicking a polycrystalline structure. We observe bursts of magnetic activity resulting 
from repeated domain wall pinning due to the disorder and subsequent depinning triggered 
by the dipolar interaction between the moving films. These domain wall jumps 
result in strong energy dissipation peaks. We study how the properties of the polycrystalline structure such as grain size and strength of the disorder, along with the driving velocity and the width of the films, affect the magnetization dynamics, average energy dissipation as well as the statistical properties of the energy dissipation bursts.
\end{abstract}
\maketitle

\section{Introduction}

Crystalline structures of solids found in nature are rarely perfect, but instead contain many kinds of impurities, defects and grains. These irregularities determine the mechanical, thermal and electromagnetic properties of materials to a great extent, examples of which can be found from the production of electrical steels \cite{electricalsteel} to doping semiconductors \cite{semiconductors}. Modern fabrication techniques have made it possible to purposefully engineer materials at very small scales to have desirable micro- and macroscopic properties. 

When it comes to magnetic properties of materials, an interesting consequence of the imperfect lattice structure is the creation of energetically preferable locations in which magnetic substructures, such as domain walls and vortices, can become pinned. The pinning contributes to multiple static and dynamic attributes of the magnet, from affecting properties such as coercivity and permeability\cite{Grainsize2}, to giving rise to dynamics such as Barkhausen jumps/avalanches\cite{durinBarkhausen}, in which the domain walls inside a magnet jump from one configuration to another during a magnetization process.

Changes in magnetization, such as the aforementioned Barkhausen avalanches, incur energy losses due to various coupling mechanisms between the spin degrees of freedom and the lattice\cite{Dissipation}. Along with eddy current losses, these losses due to magnetic dynamics, including hysteretic losses due to domain wall jumps and anomalous losses, are relevant in applications where there are high-frequency alternating electromagnetic fields and/or components moving in such fields\cite{magneticmaterials}, such as in magnetic bearings\cite{Bearings}, magnetic gears\cite{Gears2} and electric motors\cite{motors}. In thin films and insulators, the hysteretic losses are particularly important due to eddy currents being largely negligible \cite{magneticmicrostructures}. When associated with motion, it is natural to call the magnetic losses ''magnetic friction''.

In this study, we investigate how a disordered polycrystalline structure and the related domain wall pinning and depinning influence the magnetic domain wall dynamics and the resulting magnetic friction between thin films in relative motion. We focus on two things: the influence of the dimensions of the films and parameters such as grain size and strength of the disorder on the average energy dissipation, and the statistics of the fluctuations in the energy dissipation due to bursts of domain wall motion in the system.

The paper is structured as follows. In Sec. II, we go through the theoretical background of effects of grain size on the properties of polycrystalline magnets and domain wall motion in a disordered medium. Sec. III explains our micromagnetic simulation scenario and the relevant details regarding the interaction of magnets and energy dissipation. The results of the simulations are provided in Sec. IV, and the conclusions of this study are elaborated on in Sec. V.

\section{Domain structure and dynamics in polycrystalline magnets}

Between perfectly monocrystalline structure and completely disordered (amorphous) structure are polycrystalline solids, a common form of structure found in e.g. metals and ice. Polycrystalline solids consist of multiple single-crystal grains (crystallites) with more or less random sizes and crystallographic orientations, determined by conditions in which the solid is formed. In magnetic materials, the individual grains influence the total domain structure of the magnet, their contribution determined by the orientation of grain surface and grain boundaries relative to the orientation of the easy anisotropy axis/axes and the interaction between nearby grains \cite{BarkhausenBook}.

The magnetic properties of a ferromagnet are greatly affected by the degree of structural order, or crystallinity, of the material. It has been found, for example, that due to domain wall pinning at grain boundaries, both the coercivity and remanent magnetization of nanocrystalline magnets can be tuned by altering the grain size \cite{Grainsize1}. The maximum of these properties is attained when the grain size equals the typical size of magnetic nanostructures such as domain walls, leading to strong pinning and thus domains that resist changes of size and shape by an applied external field. 

The grain size dependence is mostly the result of the competition of magnetocrystalline anisotropy and exchange interaction defining how strongly domain walls become pinned at the grain boundaries. When the grains are smaller than the typical width of domain walls in the material, the exchange interaction prevents the magnetization from completely aligning into the preferred magnetization direction of each grain, averaging the pinning disorder over multiple grains and thus lowering the effective anisotropy. Grains approximately the size of a domain wall strike a balance between following the anisotropy direction and having a slowly varying magnetization, thus achieving the strongest pinning effect. With further increasing grain size the possible pinning volume decreases, an extreme example being a single crystal magnet with no grain boundaries, in which a domain wall can in principle move freely. In larger grains, the exchange energy also plays a decreasing role in the magnetization reversal, so that the magnetization in each grain can be switched more easily.  \cite{nanomats}

\subsection{Domain wall jumps in disordered media}

A consequence of domain wall pinning is that the domain wall motion during magnetization processes of ferromagnets is not continuous, but consists of periods of inactivity followed by short bursts of movement. The magnetic dynamics is thus dominated by intermittent domain wall jumps, the character of which depends on the strength of the field driving the magnetization.  

Close to the depinning field strength $H_\mathrm{d}$ where the domain walls become completely unpinned, one encounters the Barkhausen effect\cite{BarkhausenBook}, in which the domain wall motion is dominated by large-scale avalanches across the system. The size distribution of Barkhausen jumps or avalanches has been found to contain universal characteristics similar to many other forms of crackling noise, such as earthquakes\cite{earthquakes} and microfractures\cite{fracture}. The size distribution $P(S)$ typically following a power law \mbox{$P(S) \propto S^{-\tau_\mathrm{S}}$}, with well-defined exponents $\tau_\mathrm{S}$ for a wide range of avalanche sizes, suggesting critical behavior \cite{BarkhausenExponents}. Due to interest in the statistics of critical phenomena and avalanche dynamics in disordered systems, Barkhausen noise has been quite extensively studied, both in \mbox{3-dimensional} magnets \cite{3dbarkhausen1} and thin films \cite{toukobarkhausen, Barkhausen1, Barkhausen2, Barkhausen4}. As the domain walls tend to get pinned at impurities and grain boundaries in the material, the Barkhausen signal during a magnetization process can potentially serve as a measure of probing e.g. the grain size of a ferromagnetic material \cite{BarkhausenGrain}.

Another class of domain wall motion in disordered media, taking place at field strengths below the depinning field, is the domain wall creep regime. In this regime, small segments of the domain wall undergo motion approximately independently due to thermal activation \cite{creepexp}. Studied both experimentally \cite{Creepavas1} and with simulations \cite{CreepSim}, the domain wall creep has been found to include avalanches that obey slightly different scaling than the Barkhausen avalanches at the depinning threshold \cite{creepnumer, Creepavas2}. The domain wall roughness and avalanche statistics in the creep regime have generally been found to follow the theory of an elastic interface in a random pinning landscape \cite{creepexp, creepstats}.

Compared to the aforementioned types of driven domain wall motion, where driving is accomplished by an external field and thermal effects, in our system the domain wall motion in one thin film is instigated due to the interaction with the stray field of the other film. The change in magnetization in one film affects its own stray field, further changing the response of the other magnet. Additionally, in our films the domain walls are confined to a much smaller space with a relatively small number of grains. Our interest lies in if and how these differences affect the size and duration statistics of the domain wall jumps.

\section{Simulation setup}

We simulate polycrystalline thin films in relative motion using micromagnetic simulations. Our simulation setup consists of two thin films, the upper of which is driven towards the \mbox{$+x$-direction} with a constant \mbox{velocity $v$} while the lower film is held in place. The equation of motion for the moving film is solved simultaneously with magnetization dynamics, which are governed by the Landau-Lifshitz-Gilbert equation
\begin{equation}
\frac{\partial \textbf{m}}{\partial t}=-\frac{\gamma_0}{1+\alpha^2}\big(\mathbf{H}_{\mathrm{eff}} \times \mathbf{m} + 
\alpha \mathbf{m} \times (\textbf{m} \times \mathbf{H}_{\mathrm{eff}})\big),
\label{EQLLGNumerical}
\end{equation}
where \mbox{$\gamma_0$ is} the product of electron gyromagnetic ratio $\gamma$ and the permeability of vacuum $\mu_0$, \mbox{$\mathbf{m}$ is} the normalized magnetization vector, \mbox{$\mathbf{H}_\mathrm{eff}$ is} the effective field and \mbox{$\alpha$ is} the phenomenological Gilbert damping constant. The effective field takes into account the exchange interaction, magnetocrystalline anisotropy, the demagnetizing field and the external field, the last of which was absent in the simulations of this work. We use micromagnetic solver Mumax3\cite{vansteenkiste2014design} augmented with our smooth motion package\cite{extension} to simultaneously solve the motion and magnetization dynamics. 

\begin{figure}[t!]
\leavevmode
\includegraphics[trim=0cm 0cm 0cm 0cm, clip=true,width=1.0\columnwidth]{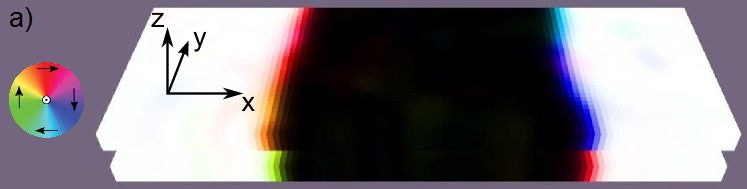}
\includegraphics[trim=0cm 0cm 0cm 0cm, clip=true,width=0.72\columnwidth]{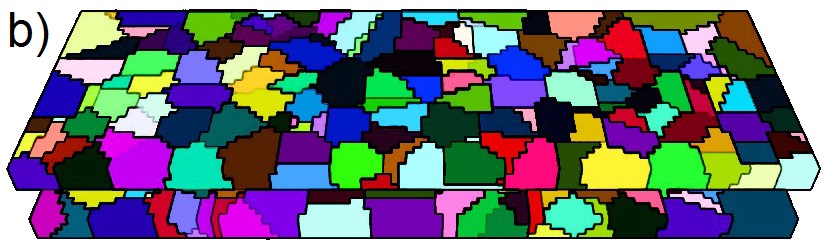}\includegraphics[trim=6cm 4.3cm 6cm 3.0cm, clip=true,width=0.28\columnwidth]{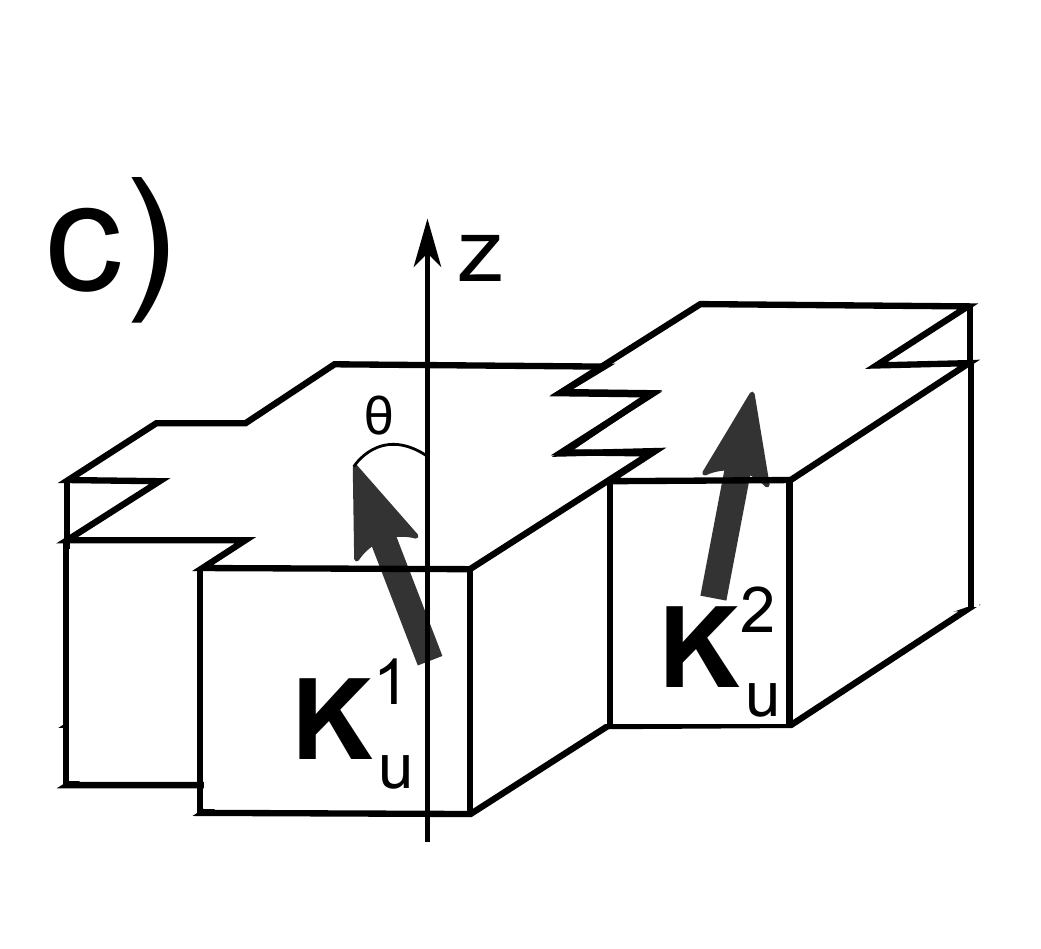}
\caption{\textbf{a)} The two films and the initial magnetization configuration (periodic images not shown). The color wheel shows the orientation of the magnetization in-plane, whereas black and white correspond to $-z$ and $+z$-directions, respectively. \textbf{b)} The schematic depiction of the Voronoi tessellated grains of the two films. \textbf{c)} The realization of the disorder: anisotropy vectors with randomized deviations from the $z-$axis in a pair of grains.}
\label{FIGSimSetup}
\end{figure}

The micromagnetic parameters were chosen to represent a hard, uniaxial material, with \mbox{CoCrPt-like}\cite{CoCrPt} parameters \mbox{$M_\mathrm{sat} = 300$ kA/m}, \mbox{$A_\mathrm{ex} = 10^{-12}$ J/m}, \mbox{$K_\mathrm{u} = 200$ kJ/m$^3$} \mbox{and $\alpha = 0.01$}. The uniaxial anisotropy easy axis points along the \mbox{$z-$axis} (out of film plane). The films were \mbox{20 nm} thick and \mbox{424 nm} long (though the periodicity makes them effectively infinite), with the film width being one of the studied variables in this study. The simulation domain was discretized into \mbox{4 nm  $\times$ 4 nm $\times$ 4 nm} cells, and the distance between the films was set to 5 cells (\mbox{20 nm}), so that they can be considered to interact only via the demagnetizing field. We ignore thermal effects, running the simulations in \mbox{0 K} temperature.

The initial magnetization is a simple structure of two domains, starting aligned in both films, with one domain having \mbox{$+z$-directional} magnetization and the other domain having \mbox{$-z$-directional} magnetization \mbox{(Fig.~\ref{FIGSimSetup} \textbf{a})}. The simulation volume is periodic in the driving direction, resulting in two Bloch domain walls between the two domains.

In this paper we focus on polycrystallinity, ignoring other forms of lattice irregularities that could cause pinning. The films were made polycrystalline by dividing them into grains with Voronoi tessellation\cite{disorder} \mbox{(Fig.~\ref{FIGSimSetup} \textbf{b})}. As the theoretical estimate for the domain wall width using the aforementioned material parameters is approximately \mbox{$l_\mathrm{dw} = \pi\sqrt{A_\mathrm{ex}/K_\mathrm{u}} \approx 22$ nm}, we simulate the films with three average grain diameters $\langle D \rangle$, \mbox{10 nm}, \mbox{20 nm} and \mbox{40 nm}, taking into account the theoretical consideration of strongest pinning being found with grain size roughly equal to the domain wall width. 
 
There are multiple ways to realize the disorder in the tessellated films, e.g. weakening the exchange interaction between grains or changing material parameters such as $M_\mathrm{sat}$ within the individual grains. In our system, the magnetocrystalline anisotropy is the dominating energy term, and thus we chose to simulate the disorder by deviating the direction of the anisotropy vector from the $z-$axis by a random amount in each grain, with \mbox{$x-$ and $y-$components} $\Delta x$ and $\Delta y$ drawn from the normal distribution with mean 0 and equal standard deviations \mbox{$\sigma_x=\sigma_y=\sigma$}. The length of the anisotropy vector is 1 in the $z-$direction, and thus together the deviations form an angle $\theta = \tan^{-1}(\sqrt{\Delta x^2+\Delta y^2})$ from the $z-$axis (Fig.~\ref{FIGSimSetup} \textbf{c}). After the randomized deviation, the vector is normalized to unit length again to keep the magnitude of the anisotropy constant.

\subsection{The interaction of magnets and magnetic losses}

In our simulation scenario with two films, the domain dynamics of the films couple via the demagnetizing fields of the mutually changing magnetization. Initially, the magnets relax into an equilibrium in which the total energy of the system is minimized, usually meaning aligned domains in films close together. As the motion of the driven upper film moves the domains within it, the \mbox{$+z$ and $-z$ domains} of the stationary and moving film become misaligned, increasing stray field energy. The resulting dynamics depends on how strongly the domain walls are pinned in the films.

\begin{figure}[t!]
\leavevmode
\includegraphics[trim=0.5cm 1.3cm 0.5cm 1.5cm, clip=true,width=1.0\columnwidth]{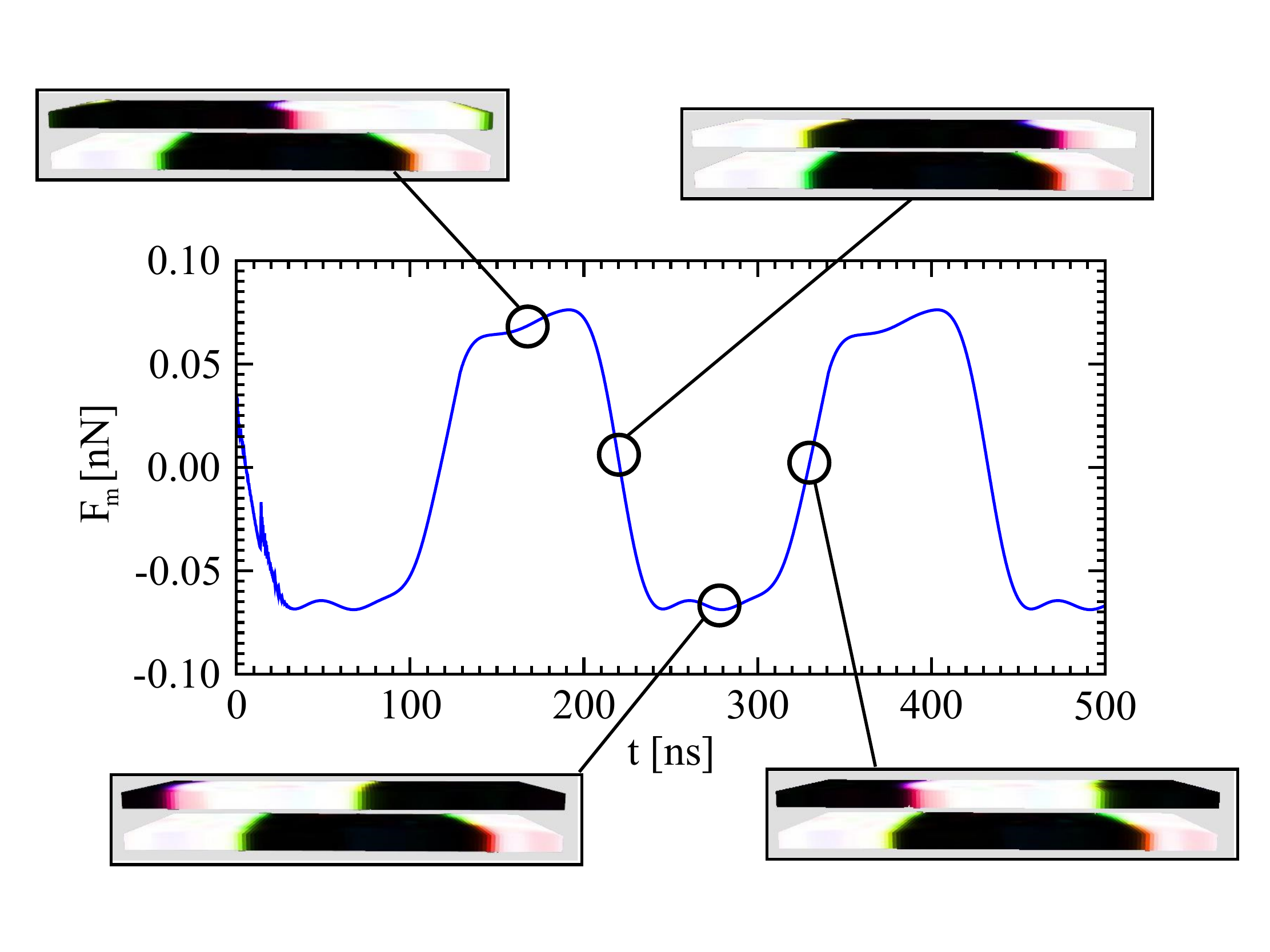}
\caption{In a completely pinned system, the energy is periodically stored and released due to the domains aligning and misaligning, resulting in an oscillating magnetic force being exerted on the films. Due to the pinning, there's negligible dissipation, and the magnetic force between the films is zero on average.}
\label{FIGEnergyStorage}
\end{figure} 

In the case of weak pinning, the stray field of the magnets exceed the depinning field, and the domain walls in the films tend to match positions. Depending on the random grain pattern and disorder strength in the grains, this means that the domain walls either stay still on average (stationary film has stronger pinning) or move towards the driving direction at a fraction of the driving velocity (moving film has stronger pinning). Contrariwise, if the pinning is strong, the stray field is not enough to depin the domain wall(s), in which case the domain walls stay misaligned when the moving film is driven forward, increasing stray field energy. In this case, the resulting energy gradient tries to drive the displaced film back to it's original location. The magnetic force acting on the moving film is determined by
\begin{equation}
\mathbf{F}_\mathrm{m} = \mu_0V_\mathrm{cell}M_\mathrm{sat}\sum_{i\in \{u\}}\nabla(\mathbf{m}^\mathrm{i}\cdot\mathbf{H}_\mathrm{l}^\mathrm{i}),
\label{EQForce}
\end{equation}
where \mbox{$\mu_0$ is} the permeability of vacuum, \mbox{$V_\mathrm{cell}$} is the discretization cell volume, \mbox{$M_\mathrm{sat}$} is the saturation magnetization, \mbox{$\{u\}$ denotes} the discretization cells belonging to the upper (moving) film, \mbox{$\mathbf{m}_\mathrm{i}$ is} the magnetic moment vector in discretization \mbox{cell $i$} and \mbox{${H}^\mathrm{i}_\mathrm{l}$ is} the demagnetizing field of the lower (stationary) film acting on \mbox{cell $i$}. In a completely pinned system the magnetic force oscillates indefinitely (Fig.~\ref{FIGEnergyStorage}) due to the periodic misalignment and realignment of the up and down domains.

Though \mbox{$\mathbf{F}_\mathrm{m}$ resists} the motion of the moving film, it is not dissipative per se, and thus is not contributing to the magnetic losses and thus magnetic friction. The magnetic losses mainly originate from the pinning and depinning of domain walls, i.e. hysteresis losses \cite{Magneticlosses}. In the micromagnetic picture, the energy dissipation comes from the relaxation of the magnetic moments according to the LLG equation after a domain wall jump. The equation for the power dissipation can be derived with the help of the LLG equation\cite{llgFriction1},

\begin{equation}
P = \frac{ \alpha\mu_0 \gamma M_\mathrm{sat} V_\mathrm{cell}}{1+\alpha^2}\sum_{i=1}^N\big(\mathbf{m}_\mathrm{i}\times \mathbf{H}_\mathrm{eff,i}\big)^2,
\label{EQDissipation}
\end{equation}
where $\mathbf{m}_\mathrm{i}$ and $\mathbf{H}_\mathrm{eff,i}$ denote the local magnetization and effective field in discretization cell $i$, respectively, and $N$ is the total number of discretization cells.

The average friction force can be calculated from the power dissipation divided by the velocity of the moving film $\langle F_\mathrm{fric}\rangle = \langle P\rangle /v$. Unless otherwise stated, we use driving velocity \mbox{$v = 1$ m/s}, so that the average dissipation power indicates also the magnitude of the average friction force.

\section{Results and discussion}

\subsection{Domain dynamics and energy dissipation} 

\begin{figure}[t!]
\leavevmode
\includegraphics[trim=0cm 0cm 0cm 0cm, clip=true,width=1.0\columnwidth]{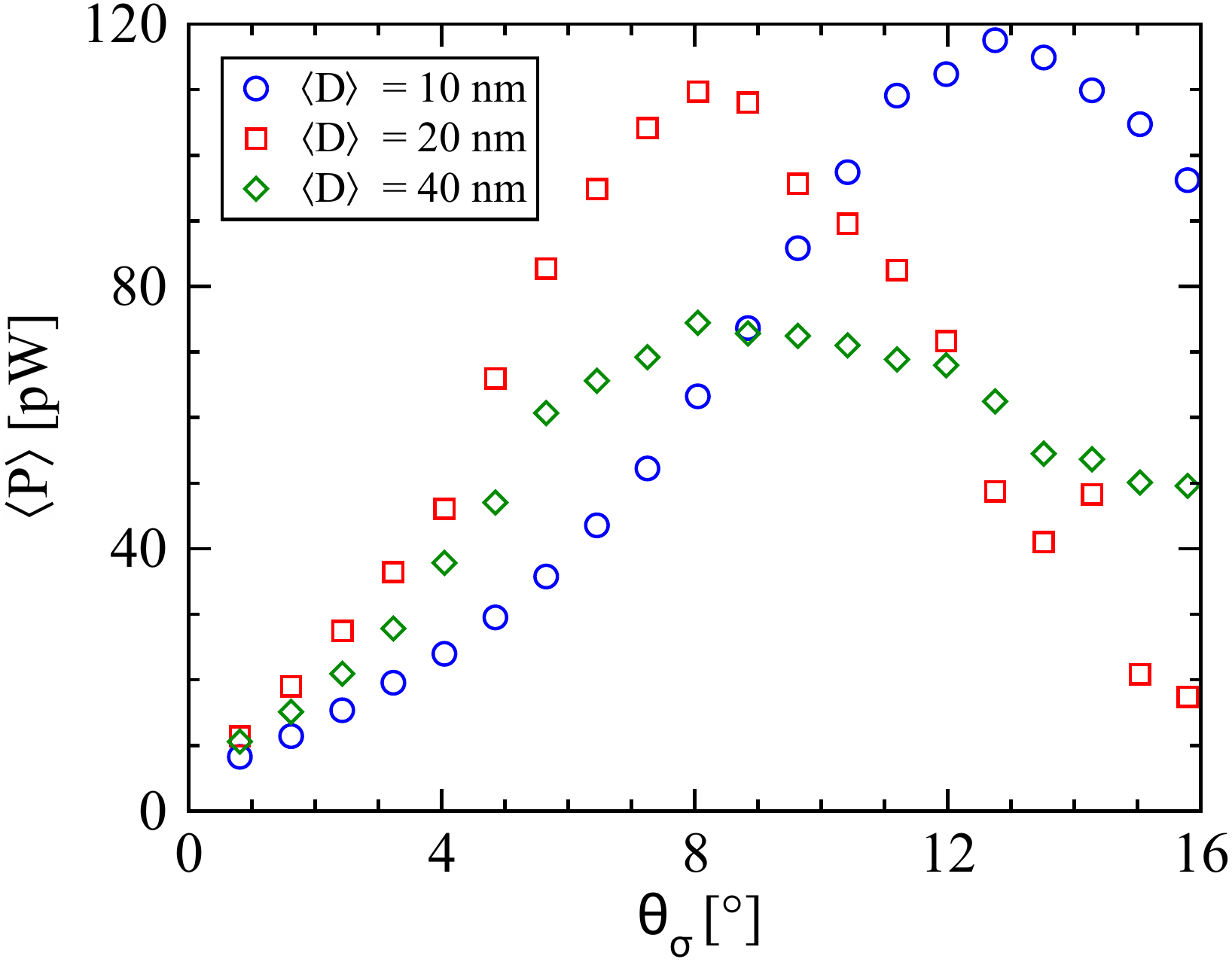}
\caption{The magnetic losses as a function of $\theta_\sigma$, the maximum angle deviation from the $z$-axis for the anisotropy vector, for the three different grain sizes. The results for the largest values of $\theta_\sigma$ are more noisy due to the pinning depending strongly on the random grain configuration.}
\label{FIGEnergyDissipation1}
\end{figure}

\begin{figure}[t!]
\leavevmode
\includegraphics[trim=0cm 0cm 0cm 0cm, clip=true,width=1.0\columnwidth]{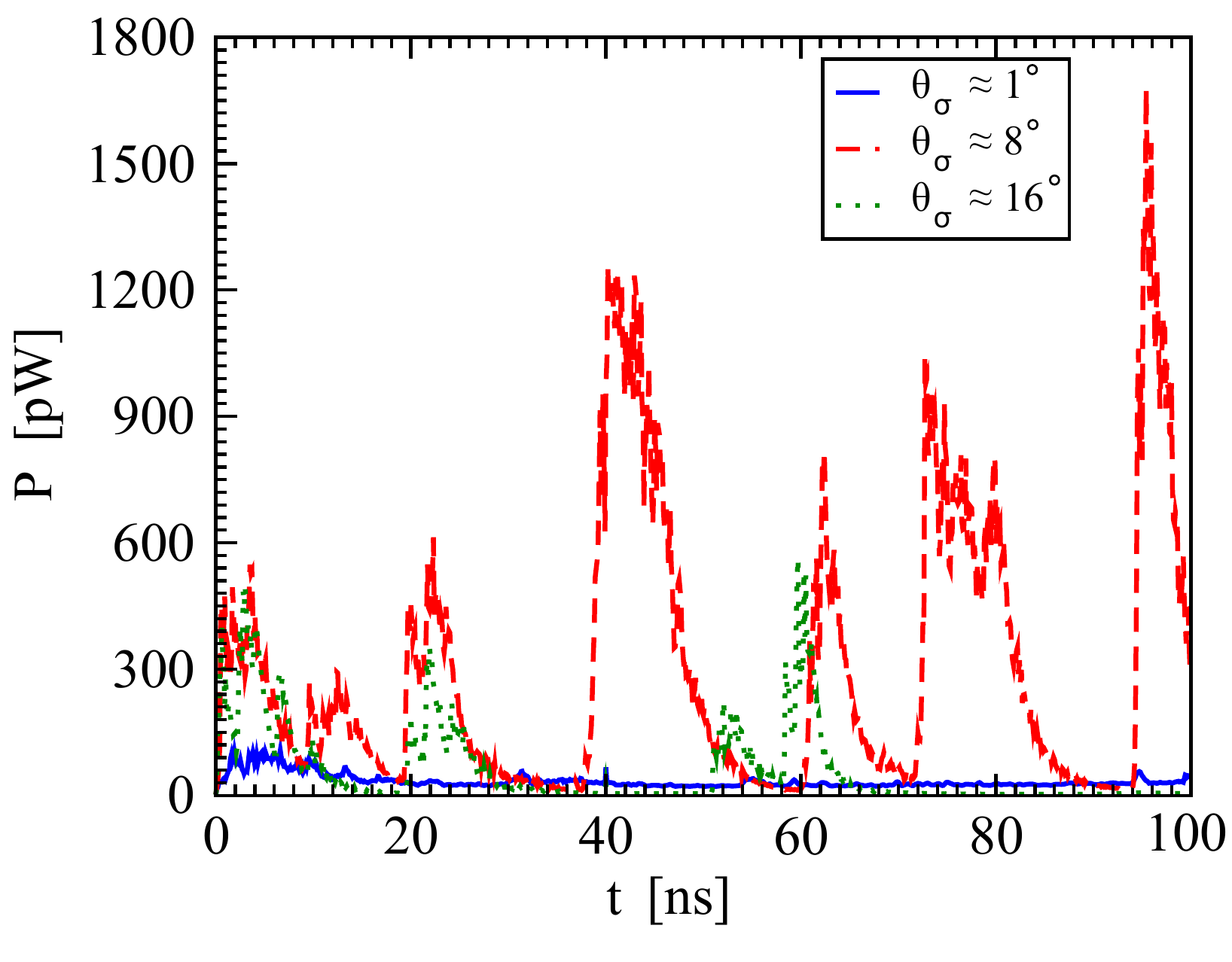}
\caption{The energy dissipation for the first 100 ns for three different disorder strengths $\theta_\mathrm{\sigma}$, showing the largest and most frequent avalanches for a system with medium strength disorder. The system is 600 nm wide with average grain diameter $\langle D\rangle = 20$ nm.}
\label{FIGdissipations}
\end{figure}

We first charted the domain wall dynamics qualitatively using a film of small width \mbox{($w = 140$ nm)} and average grain size \mbox{$\langle D \rangle = 20$ nm}, varying the standard deviation $\sigma$ to examine how the randomness in the anisotropy vector deviations affected the domain wall dynamics. For convenience, we use a single measure for the strength of the disorder, \mbox{$\theta_\sigma$, defined} as the angle in which both \mbox{$\Delta x$ and} \mbox{$\Delta y$ are} equal to one standard deviation \mbox{$\sigma$},
\[
\theta_\sigma = \tan^{-1}(\sqrt{\sigma^2+\sigma^2}) = \tan^{-1}(\sqrt{2}\sigma).
\]
We found that for low values of disorder, $\theta_\sigma < 2^\circ$, the pinning is weak, resulting in mostly smooth changes in magnetization, with the domain walls moving almost continously in response to the driving. In the regime of larger deviations ($2^\circ < \theta_\sigma < 16^\circ$), the dynamics consists of avalanche-like bursts of motion of the domain walls, interspersed with periods of negligible activity due to the domain walls being pinned. The extent of pinning and the sizes of individual avalanches depend on the strength of the pinning, with a further increase in $\theta_\sigma$ typically resulting in the domain walls not depinning at all after some initial reconfiguration. In this case the magnetization is completeley rigid, eliminating magnetic losses and thus magnetic friction. 

Based on these qualitative observations, we simulated $\theta_\sigma$ values from $1^\circ$ to $16^\circ$ for the three different grain sizes (Fig.~\ref{FIGEnergyDissipation1}). For these simulations, we used 600 nm wide films so that the films can fit a large number of grains along the domain wall, mitigating the random noise in the results. The results are also averaged over several random realizations of the grain structure. Examples of the dissipation signal in single simulations with varying strengths of disorder are shown in Fig.~\ref{FIGdissipations}.

At small disorder strengths, there's very little domain wall pinning with all grain sizes, and thus the grain size has a negligible effect on the average energy dissipation, the magnitude of which was roughly $\langle P\rangle = 10 - 30$ pW, which is still quite high compared to purely monocrystalline systems \cite{omapreprint}. When the domain walls begin to pin more strongly, the effect of grain size becomes more pronounced. As expected from the initial simulations, for the average grain diameter \mbox{$\langle D\rangle=20$ nm}, the lowest and highest $\theta_\sigma$ values both show diminishing magnetic losses, due to either having practically no pinning at all ($\theta_\sigma \leq 1^\circ$) or mostly pinned domain walls ($\theta_\sigma \geq 16^\circ$). As can be seen from Fig.~\ref{FIGdissipations}, the few avalanches that occur at the strongest disorder are sharp and short, with long downtimes in between. The peak dissipation, where the average dissipation power and thus friction force is roughly an order of magnitude stronger, was found to lie at roughly in the middle of the two extremes, $\theta_\sigma = 7^\circ-9^\circ$. In this regime, the magnetization dynamics is governed by individual parts of the domain wall undergoing intermittent wide avalanches. The average friction force in this regime, $\langle F_\mathrm{fric}\rangle \approx 0.1$ nN, is quite high compared to forces usually encountered in non-contact friction \cite{tribobook}.

\begin{figure}[t!]
\leavevmode
\includegraphics[trim=0cm 0cm 0cm 0cm, clip=true,width=1.0\columnwidth]{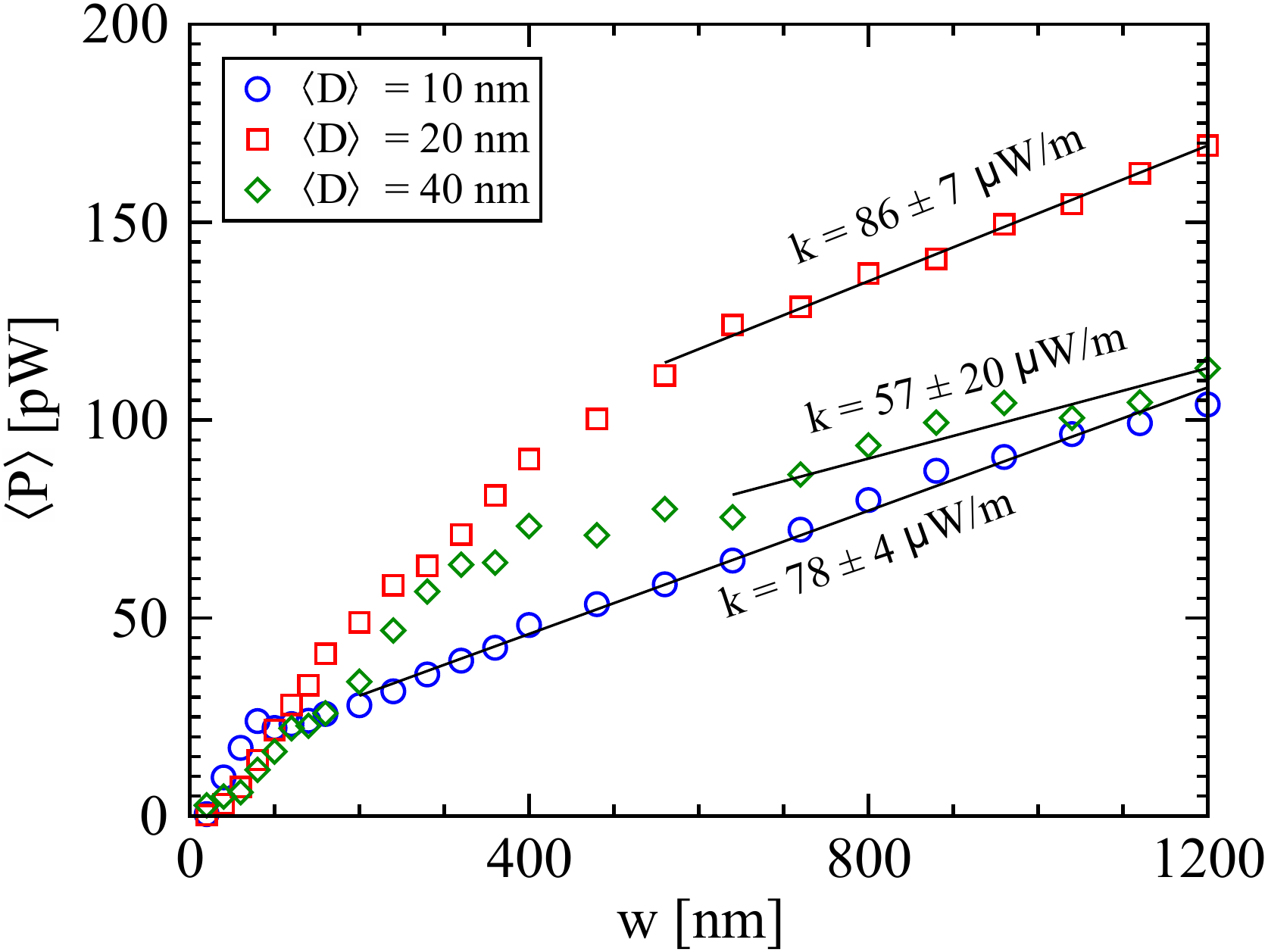}
\caption{The average dissipation power as a function of film width $w$ with \mbox{$\theta_\sigma = 8 ^\circ$} and driving velocity $v = 1$ m/s. For $\langle D\rangle = 10$ nm and $\langle D \rangle = 20$ nm, the curve becomes linear when the size of the film exceeds roughly 15 times the average grain size. The result is likely similar for largest grain size, but due to noise the linearity isn't as clear.}
\label{FIGEnergyDissipation2}
\end{figure}

For \mbox{$\langle D\rangle = 10$ nm}, we observe the previously-discussed averaging\cite{nanomats} effect due to having smaller grain size than the domain wall width, and thus lowered effective anisotropy. While not exactly half, the measured dissipation for \mbox{small $\theta_\sigma$} is much lower, and the dissipation peak is found at almost double the angle deviation compared to \mbox{$\langle D\rangle = 20$ nm}, since at this point the larger deviations in the anisotropy direction balance out the averaging due to grain size. The overall low dissipation with the largest grain size $\langle D\rangle = 40$ nm is likely a combination of the easier switching of each grain and the lowered pinning volume due to the films still not being large enough to accommodate many grains along the direction of the domain wall ($y-$direction). The dissipation for the largest grain size also display the most noise, since the small number of grains leads to the results depending more on the grain configuration.

\begin{figure}[t!]
\leavevmode
\includegraphics[trim=0cm 0cm 0cm 0cm, clip=true,width=1.0\columnwidth]{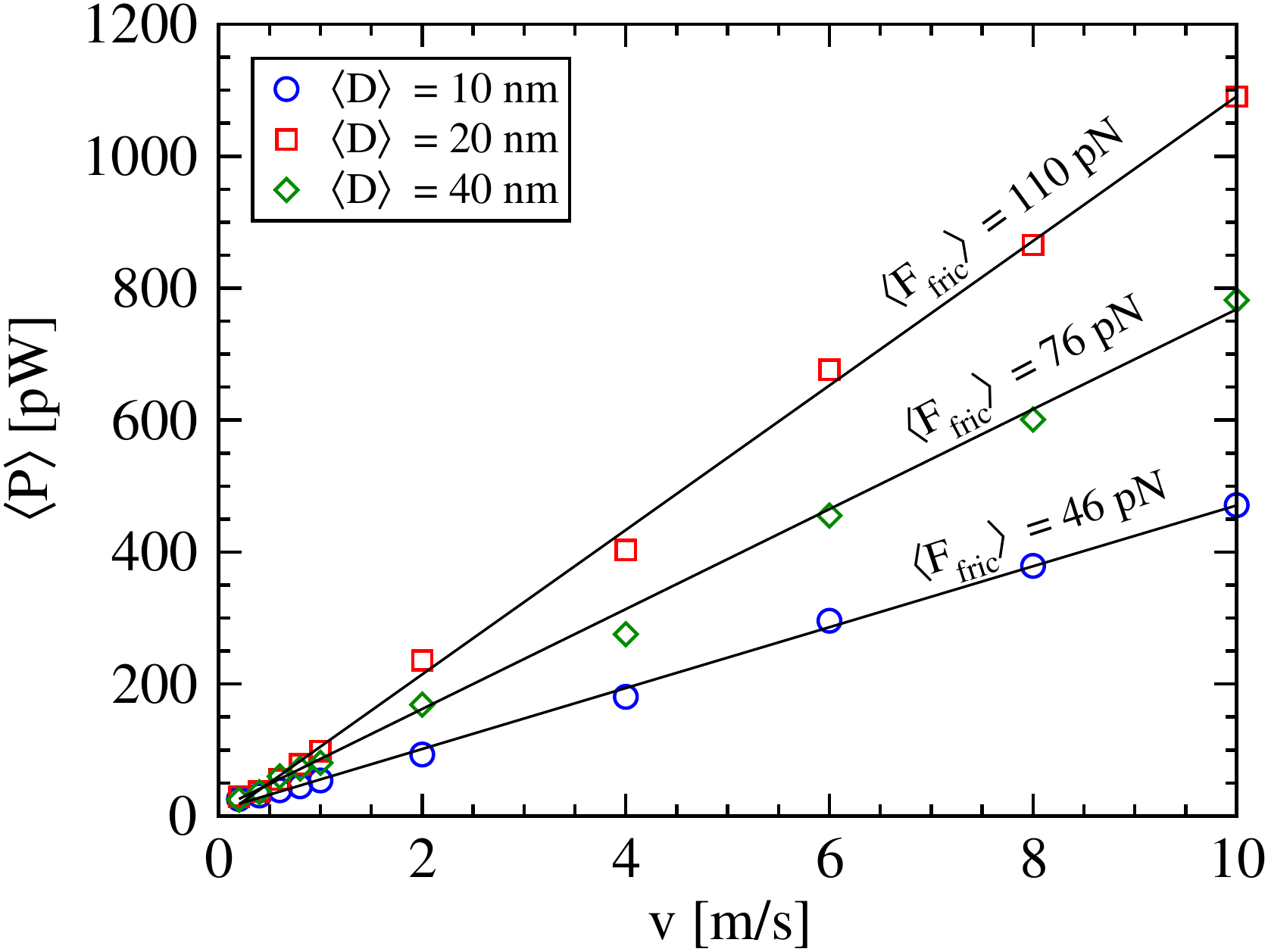}
\caption{The average dissipation power as a function of velocity with \mbox{$\theta_\sigma = 8 ^\circ$} and $w = 600$ nm. The relationship is linear for all grain sizes, implying that the average friction  force is independent of velocity in this range.}
\label{FIGVelocity}
\end{figure}

Being interested in how width of the film $w$ compared to the average grain size affects the dissipation, we varied the width of the films in the $y-$direction from \mbox{20 nm up} to \mbox{1.2 $\mu$m}. The energy dissipation of the system as a function of the width is depicted in Fig.~\ref{FIGEnergyDissipation2}. In these simulations, we used $\theta_\sigma = 8^\circ$, and since the angle is not at the peak of dissipation for \mbox{10 nm and 40 nm grains}, the magnetic losses with these grain sizes tend to be consistently below that of the $\langle D\rangle=$ 20 nm films for larger values of $w$. When the film can accommodate only a few grains along the domain wall, as is the case for the smallest widths, the dissipation is quite random, likely depending strongly on the random pattern of grains and anisotropy vector angles. This also causes the smallest grain size $\langle D \rangle = 10$ nm to have the strongest dissipation initially, mostly because having multiple grains in $y-$direction and relatively weaker pinning makes it more likely for the domain wall to actually depin and dissipate energy, whereas smaller number of larger grains have a higher chance to cause complete pinning of the domain wall. For $\langle D\rangle = 40$ nm in particular, the films on the smaller side contain only one or two grains along the domain wall.

After a certain point, seemingly \mbox{about 14 - 20 grains} fitting in the width of the film, the dissipation becomes more regular and starts to grow approximately linearly with size. This makes sense, as the domain wall jumps occurring in the medium disorder regime are mainly responsible for the dissipation, and the domain wall extends across the film in the $y-$direction and thus scales directly with the film width. The slopes of the linear portions for each grain are very close within error margins ($k \approx 75 - 85~\mu$W/m), though with the largest grain size $\langle D\rangle =$ 40 nm, the curve has significantly more noise and the linearity seems to appear quite late. In the linear regime, only parts of the domain walls usually depin and move at a time, whereas in low width films, the small amount of grains tends to increase the likelihood of film-wide jumps, in which the domain walls pin and depin completely at once. This results in the nonlinear (and noisy) initial growth of the dissipation power with size.

Finally, we investigated the effect of the driving velocity on the average power dissipation. It turns out that for velocities ranging between 0.2 m/s and 10 m/s, the relationship between the velocity and dissipation power is linear for all grain sizes (Fig.~\ref{FIGVelocity}), meaning that the average friction force $\langle F_\mathrm{fric} \rangle = \langle P\rangle/v$ is independent of velocity in this range. The result is analogous to hysteresis losses in hysteresis loop experiments, in which the power dissipation has been found linearly dependent on the applied field frequency \cite{bertottilosses}. The reason for the linear dependence is that the increase in frequency reduces downtime between individual domain wall jumps without significantly affecting the jumps themselves. In our setup, an increase in driving velocity has a similar effect. High velocities have the magnetic film in a near constant state of excitation due to new avalanches starting before previous ones have stopped. 

Velocities significantly exceeding 10 m/s can result in the domain pattern breaking down, resulting in single-domain films and negligible dissipation. Thus the observed linear relationship can break at high velocity. Going to much lower velocities is impractical with micromagnetic simulations as the simulation times grow considerably and the results become noisy due to only a few avalanches occurring, requiring more averaging.

\subsection{Domain wall roughness and avalanche statistics}

To acquire sufficient statistics about the domain wall avalanches, a large number of relatively long simulations is required. As such, we study the avalanche statistics using average grain size of \mbox{$\langle D\rangle = 20$ nm} with $\theta_\sigma = 8^\circ$, since these parameters results in most avalanches based on the earlier simulations. We limited the study to the effects of film width and velocity on the avalanche statistics, using three different values for the width, \mbox{200 nm}, \mbox{400 nm} and \mbox{800 nm}, and three velocities, \mbox{1 m/s}, \mbox{3 m/s} and \mbox{5 m/s}. 

\begin{figure}[t!]
\leavevmode
\includegraphics[trim=0cm 0cm 0cm 0cm, clip=true,width=1.0\columnwidth]{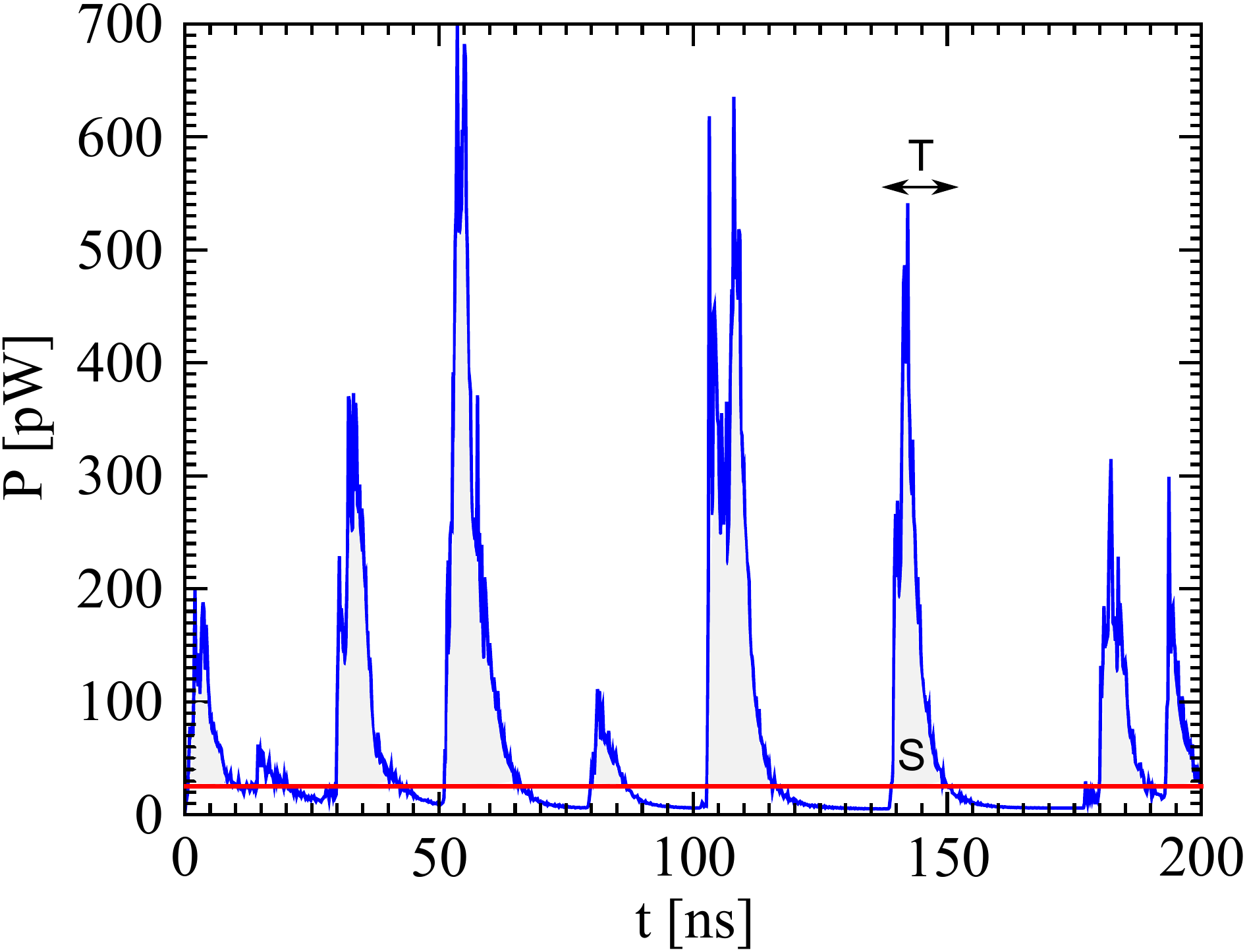}
\caption{An example of a power dissipation signal found in this study, with multiple avalanche-like bursts of magnetic activity. The size (S) of an avalanche is defined as the total amount of energy dissipated and the duration (T) as the total time the signal stays over the avalanche threshold (red line). This particular case was simulated with film width \mbox{$w=400$ nm} and driving velocity \mbox{$v=1$ m/s}.}
\label{FIGTheory}
\end{figure}

\begin{figure}[t!]
\leavevmode
\includegraphics[trim=0cm 0cm 0cm 0cm, clip=true,width=1.0\columnwidth]{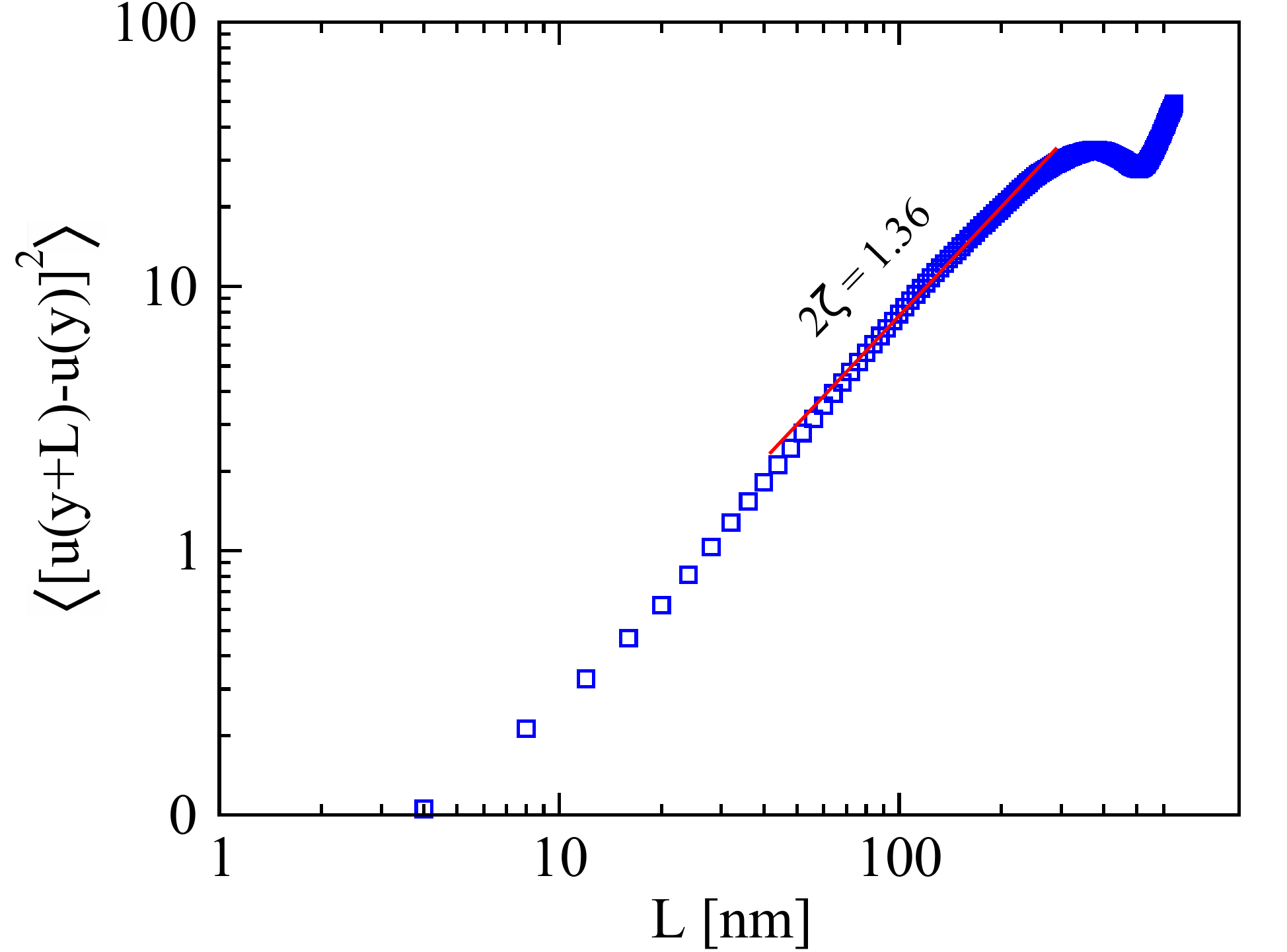}
\caption{The correlation function of domain wall displacements in a $800$ nm wide system with $\langle D\rangle = 20$ nm grains, used to find the roughness exponent $\zeta \approx 0.68\pm 0.05$.}
\label{FIGroughness}
\end{figure}

\begin{figure*}[t!]
\leavevmode
\includegraphics[trim=0cm 7.9cm 0cm 8.7cm, clip=true,width=2.0\columnwidth]{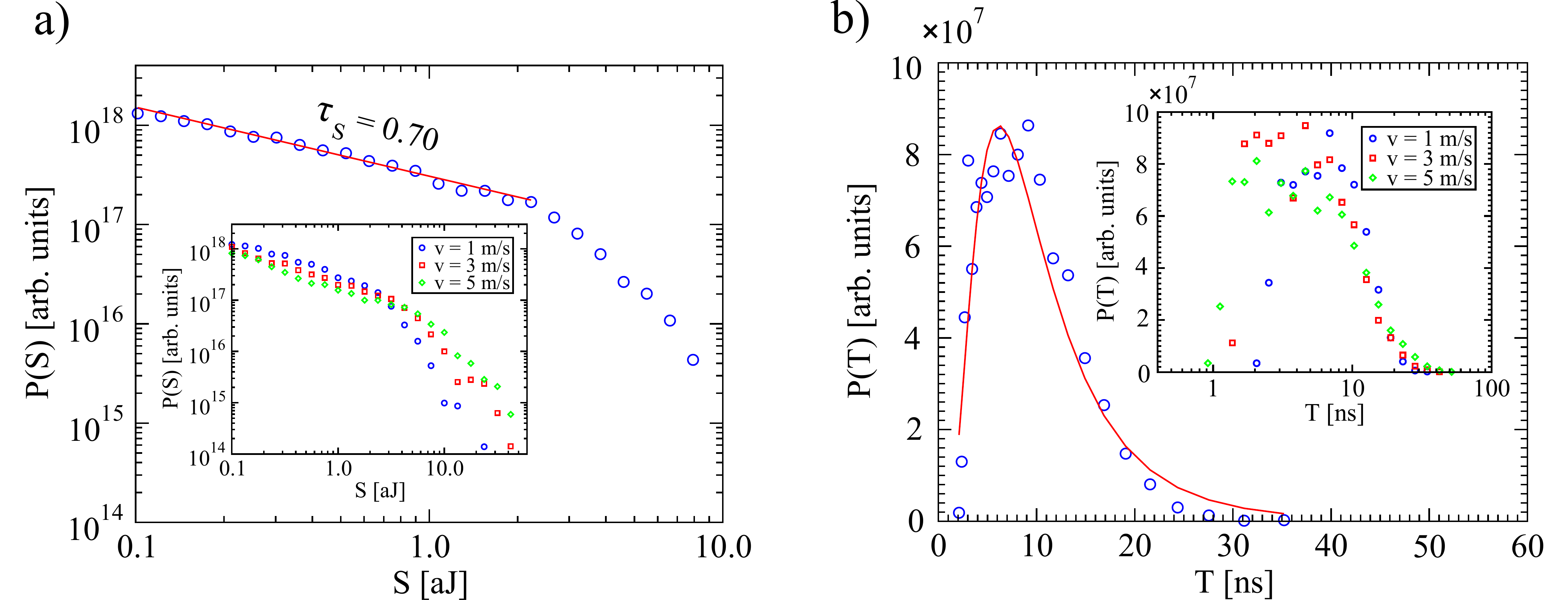}
\caption{\textbf{a)} The avalanche size distribution for a $w=800$ nm film, with a power law fitted via the maximum likelihood method. The inset shows how increasing velocity softens the cutoff due to more avalanche overlap resulting in larger avalanches. \textbf{b)} The avalanche duration seemingly follows the log-normal distribution. Larger velocities result in more broad distribution and the shifting of the peak to lower values.}
\label{FIGStats}
\end{figure*}

To capture the avalanches from the power dissipation signal $P(t)$ (Fig.~\ref{FIGTheory}), a threshold needs to be set to differentiate between an avalanche and a pinned state. Depending on the threshold, a single avalanche could be split into multiple sub-avalanches, thus changing the shape of the distribution. We found that having a threshold of \mbox{$0.3\langle P\rangle$ resulted} in a suitable amount of avalanches without incurring significant splitting. We cut off avalanches smaller than 0.1 aJ, since smaller avalanches were usually the result of the noise in the signal just momentarily crossing the avalanche threshold. 

A relevant quantity related to elastic interfaces moving in a disordered medium, such as the domain walls in our case, is the interface roughness exponent $\zeta$. Assuming non-anomalous scaling\cite{anomalous}, the roughness exponent can be found through the displacement-displacement correlation function of the domain wall displacement $u(y)$ perpendicular to the wall, $\langle [u(y+L)-u(y)]^2\rangle \propto L^{2\zeta}$, where $L$ is the distance between two points of the domain wall. Taking snapshots of a single domain wall in the upper film after 60 ns of driving, averaged over 5 random realizations of the disorder, we find $\zeta \approx 0.68\pm 0.05$ (Fig.~\ref{FIGroughness}) for the three film widths.

The avalanche size and duration distributions from simulations with film width $w = 800$ nm and driving velocity $v=1$ m/s are depicted in Fig.~\ref{FIGStats}. The size distribution resembles a power law $P(S) \propto S^{-\tau_\mathrm{S}}$ over roughly one decade $S =$ 0.1 aJ - 2 aJ, after which there's a cut-off. A likely source for the cutoff is the fact that the stray fields of the films attempt to align the domain wall locations in the upper and lower films to minimize the stray field energy, meaning that it is difficult for the domain wall experiencing an avalanche to jump past its corresponding domain wall in the other film. The extent of the jump is thus limited by the domain wall width and the grain size in the direction perpendicular to the wall.

Fitting a power law using the maximum likelihood method\cite{powerlaw1} with the NCC toolbox \cite{powerlaw2}, we find a size exponent $\tau_\mathrm{S} \approx 0.70$. Similar exponents have been predicted for weak-field driven, thermally activated avalanches in the domain wall creep regime \cite{Creepavas1,creepnumer}. The avalanche size exponent for systems with short-range disorder can also be estimated theoretically from the dimensionality of the interface $d$ and the roughness,
\[
\tau_\mathrm{S} = 2-\frac{2}{d+\zeta}.
\]
In our case, the roughness and dimensionality ($d=1$) predict $\tau_\mathrm{S} \approx 0.77 - 0.84$ for the size distribution exponent, matching quite well to the simulation results.

\begin{figure*}[t!]
\leavevmode
\includegraphics[trim=0cm 7.9cm 0cm 8.1cm, clip=true,width=2.0\columnwidth]{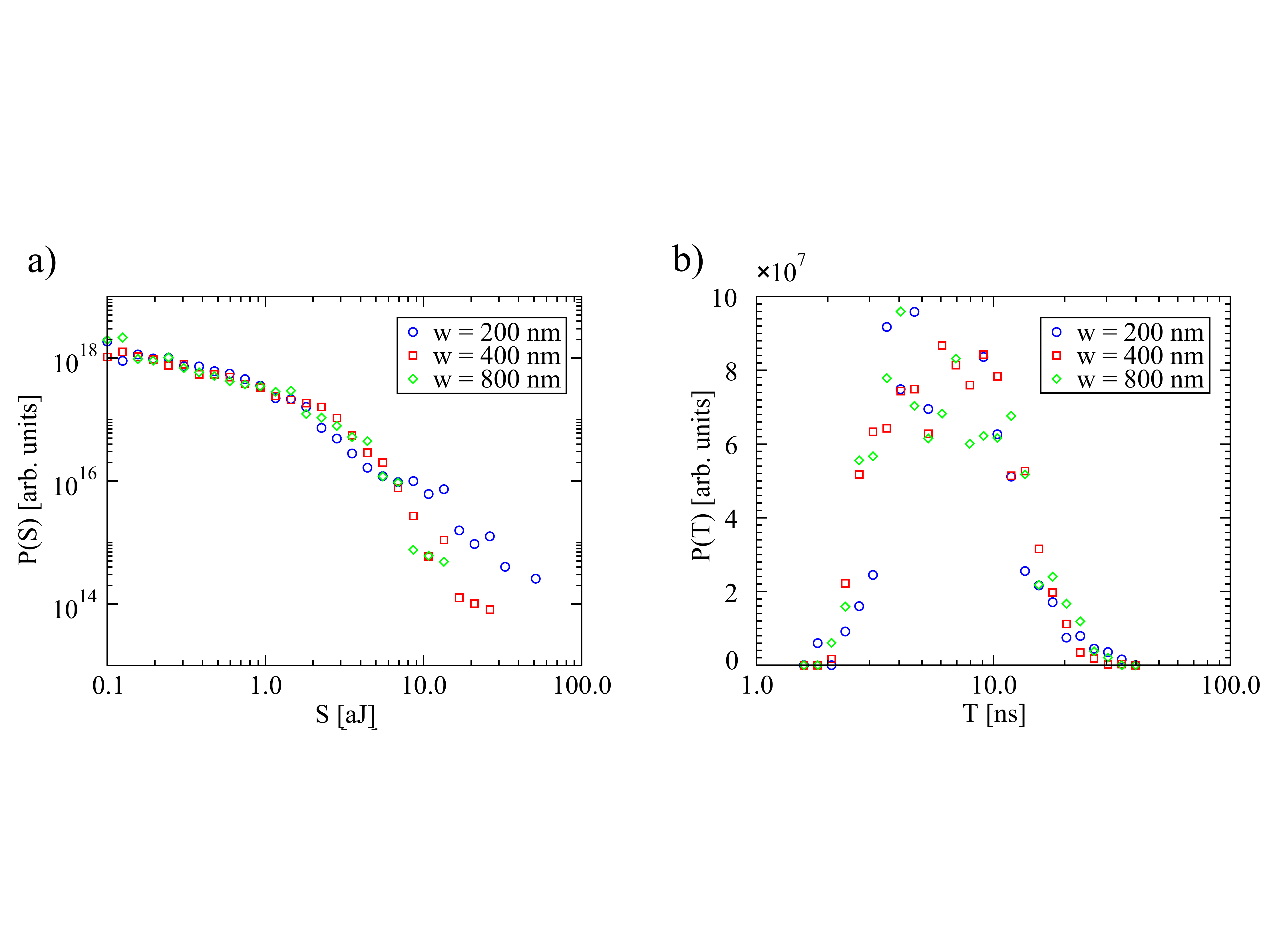}
\caption{\textbf{a)} The avalanche size distribution as a function of film width. System-wide avalanches contribute to the sizes beyond the cut-off for $w=200$ nm. \textbf{a)} The avalanche duration distribution as a function of film width. Unlike in the case of velocity, there's no visible shift in the peak of the distribution, though larger film widths display a somewhat similar widening of the distribution as was observed at higher velocities.}
\label{FIGStats2}
\end{figure*}

Interestingly, the avalanche duration distribution in our simulations seemingly follows a log-normal distribution instead of a power law (Fig.~\ref{FIGStats} \textbf{b}). The best fit to the results gives parameter values $\sigma = 0.6182$ and $\mu = -18.52$ with a mean avalanche duration $\langle T\rangle = \exp(\mu+\sigma^2/2) \approx 11$ ns. A possible reason for this form of the distribution is similar to the cut-off of the size distribution, in that the time in which the avalanche relaxes is roughly independent of the lateral size of the avalanche, and thus the duration of the jumps is mainly determined of the forward motion of the domain walls which is limited by the stray fields of the films. 

In a similar fashion to what was observed for the average energy dissipation, increasing the driving velocity reduces the downtime between individual avalanches, though the avalanches themselves are not strongly affected. However, distinguishing between individual avalanches becomes a challenge, as new avalanches tend to start before the system has relaxed. Thus the velocity has a visible effect on the avalanche size and duration distributions, illustrated in the insets of \mbox{Figs.~\ref{FIGStats} \textbf{a} and \textbf{b}}. High velocities didn't significantly affect the \mbox{exponent $\tau_\mathrm{S}$}, but the increased overlap in the avalanches resulted in the cutoff having a slightly softer falloff. The duration distribution becomes broader when velocity is increased, with both short and long avalanches becoming more common, likely due to the increased noise resulting in momentary crossings of the avalanche threshold and the overlapping avalanches combining the duration of multiple individual avalanches. However, the average avalanche duration remained approximately the same.

The film width had a lesser influence on the avalanche size and duration distributions. The only discernible effects were the increase of the number of large ($S$ around the cutoff) avalanches with the smallest film width of 200 nm and a small widening of the duration distribution with size, shown in Figs.~\ref{FIGStats2} \textbf{a} and \textbf{b}. The substantial increase in avalanche sizes in the smallest system is likely explained by the film-wide avalanches still occurring relatively often at this width. It's possible that the system-wide avalanches are also distributed according to a power-law, but the limited amount of avalanche counts make this difficult to ascertain. The size distribution for 400 nm and 800 nm wide films were almost identical, showing that the avalanche size cutoff does not scale indefinitely with the system width, thus enabling the normalization of the size distribution and the observed \mbox{low $\tau_\mathrm{S} < 1$.}

\section{Conclusion}

We simulated the interaction of two polycrystalline thin films in relative motion, investigating how the average energy dissipation is influenced by the disordered structure, determined by the grain diameter and the strength of the disorder, along with external parameters such as film width and the driving velocity. We also studied the size and duration distributions of domain wall jumps in this two-film system, and how the distributions are affected by the width of the film and the driving velocity.

Our results for the average energy dissipation indicate that the magnetic losses and thus magnetic friction are at their highest when the domain walls of the system are strongly but not completely pinned, such that the strength of the stray field of the films is just enough to depin the domain walls consistently. In this regime, the average losses were roughly an order of magnitude higher than either mostly freely moving or very strongly pinned domain walls. The domain wall motion is characterized by frequent jumps, in which parts of the domain wall experience avalanches roughly independently, the peak of the dissipation depending on the combination of average grain size and the strength of the disorder. The energy dissipation was found to be linearly proportional to the sliding velocity, a result akin to hysteretic losses arising from domain wall jumps in a magnetization process. The dissipation also scales linearly with film width, provided that length scales above a certain grain size dependent width are considered.

The domain wall avalanches were observed to be initiated by the misalignment of the up and down domains in the films, which increased the stray field energy until an avalanche realigned the domain walls. The sizes of the domain wall jumps seemingly followed a power-law+cutoff distribution with a relatively small exponent, a value similar to what has been found for thermally activated domain wall creep, while the avalanche duration distribution followed a log-normal distribution, presumably due to the limited extent of the avalanches. The avalanche size exponent predicted from the roughness and dimensionality of the wall agreed quite well with the exponent obtained from the simulations. Aside from a minor increase in the number of largest avalanches, the film width and the driving velocity did not significantly alter the avalanche distributions.

Overall, our results reveal intriguing physics arising from the coupled collective dynamics of interacting domain walls, resulting in bursty magnetic non-contact friction with a relatively large magnitude. The domain wall interaction as a driving force for the avalanches produced atypical, non-critical statistics, meriting further study regarding the domain wall and disorder characteristics and their relation to the avalanche distributions.

\begin{acknowledgments}
We acknowledge the support of the Academy of 
Finland via an Academy Research Fellowship (LL, projects no. 268302 and 303749), and 
the Centres of Excellence Programme (2012-2017, project no. 251748). We acknowledge 
the computational resources provided by the Aalto University School of Science 
Science-IT project and the CSC. 
\end{acknowledgments}

\bibliography{bibl}

\end{document}